\documentclass[hidelinks,report,9pt]{article}
\usepackage[utf8]{inputenc}
\usepackage{amsmath}
\usepackage{amssymb}
\usepackage{authblk}
\usepackage{mathrsfs}
\usepackage{braket}
\usepackage[colorlinks,linkcolor={blue},citecolor={red},urlcolor={red}]{hyperref}
\usepackage{empheq}
\usepackage{soul,ulem}
\title{Emission Distribution for the quantas of Maxwell-Chern-Simon Gauge Field coupled to External Current}
\author{Tiyasa Kar}
\affil[]{Sardar Vallabhbhai National Institute of Technology,\\ Ichchhanath, Surat - Dumas Road,\\
Surat 395007, Gujarat, India.}
\date{}
\begin{document}

\maketitle
\begin{abstract}
In this paper, we have investigated the nature of  emission distribution of the Maxwell Chern Simon (MCS) Theory in 2+1 dimension. The distribution of the topologically massive quanta seems to be Poissionian in nature just like the Maxwell field theory in 3+1 dimension but with a condition, without which the distribution takes an indeterminate form when we make the coupling term approach 0.
\end{abstract}
\begin{center}
\end{center}
\section{Introduction and motivation}
The notion that the gauge invariance of a vector field theory necessarily requires the existence of only massless physical particles was discarded by Schwinger in 1962 \cite{0}  and he produced a physical model popularly known after his name as Schwinger's model or QED in 1+1 dimension. This provides room for massive "photons" to co-exist along with gauge invariance and hence provided a stimulus to the search of similar mechanisms in higher dimensions. Later, Deser, Jackiw and Templeton in their paper in 1982 \cite{3} showed that in 2+1 dimension the mass terms can appear in the presence of an additional Chern-Simons term which is of topological origin and can only be written for odd dimensions like in 2+1 dimensions.\\
Together, it is called the Maxwell-Chern-Simons (MCS) theory and the term 'topologically massive' was coined as the Chern-Simons term entitles a mass to the Maxwell field. The field Lagrangian is written as $\mathscr{L}=\mathscr{L}_M+\mathscr{L}_{CS}$. Although the Lagrangian is only quasi-gauge invariant but it generates a total divergence term under
gauge transformations (if the current is conserved) and thus leads to the gauge-invariance of the action. 
\\
It is well known that, interaction of Maxwell field with external conserved current gives rise to photon emission and the distribution of number of photons is Poissonian in nature \cite{7}. When we reduce one space dimension and study the same for planar field theory the photons become spin-less. Inclusion of a topological term indeed introduces one extra degree of freedom to the system and the quanta becomes massive. So naturally a change could be anticipated due to the difference of canonical structures of both field which arises from a different Green function for the MCS theory.  The aim of this work is to calculate the probability distribution of the emission of such massive "photons" for the above mentioned planar system which to the best of our knowledge has not been addressed before in literature . It is also shown that the infrared divergence which is present in 3+1 QED can be tamed with MCS theory due to the presence of the topological mass.  \\
Also this kind of massive theory are helpful to understand many planar condensed matter phenomena.  As pure CS theory is anyonic in nature, it is very useful to describe phenomena like fractional quantum hall effect and high $T_c$ superconductivity \cite{11} , \cite{12}. \\
The paper is organised as follows. In section 2 the polarisation vector for the MCS theory has been calculated. In scetion-3  we have thoroughly calculated the propagator followed by calculation of the S matrix and  probability distribution. In section-4 we have discussed the implications of the result obtained in the previous section.
\section{Polarisation Vector in the MCS theory }
When we formulate Maxwell theory in 2+1 space-time dimension we get only one physical degree of freedom and  photons become spin less. One way to introduce spin is to compliment the Lagrangian by an extra topological term like Chern Simon term. The term is named topological as it has no metric dependence and it does not contribute to the energy-momentum tensor.The term is given as $\quad\frac{\theta}{2}\epsilon_{\mu\nu\lambda}A_{\mu}\partial_{\nu}A_{\lambda}$\\There are several features of this term which makes it distinct from the pure Maxwell term. First, though the term is not gauge invariant, under gauge transformation the Lagrangian changes by a total time derivative. Therefore by neglecting the boundary term we can show that the corresponding Chern Simon action is gauge invariant. We can thus say that the Chern Simon term is quasi gauge invariant.\\Secondly, the Chern Simon theory can only be defined in odd dimension.\\Under parity transformation $x^1\rightarrow-x^1$ and $x^2\rightarrow-x^2$, the gauge fields trnsform as $A_1\rightarrow -A_1,\quad A_2\rightarrow A_2,\quad A_0\rightarrow A_0$. Chern Simon term therefore flips its sign. It also changes sign under time reversal.\\Thirdly, this term is first order in space-time derivatives which makes the canonical structure of this theory significantly different from that of Maxwell theory. The MCS Lagrangian in 2+1 dimension is \cite{1},\cite{3},\cite{4},\cite{11},
\begin{equation}
    \label{eq:1}
    \mathscr{L}=-\frac{1}{4}F^{\mu\nu}F_{\mu\nu}+\frac{\theta}{2}\epsilon^{\mu\nu\rho}A_{\mu}\partial_{\nu}A_{\rho}\quad;\quad\theta>0
\end{equation}
where, $\theta$ is the Chern-Simon coefficient taken to be positive. The corresponding equation of motion is
\begin{equation}
\label{eq:2}
    \partial_{\alpha}F^{\alpha\beta}-\frac{\theta}{2}\epsilon^{\mu\alpha\beta}F_{\alpha\mu}=0
\end{equation}
Imposing Lorenz gauge condition $\partial_{\alpha}A^{\alpha}=0$, this yields
\begin{equation}
\label{eq:4}
    (\partial^2\eta^{\mu\beta}-\theta \epsilon^{\mu\alpha\beta}\partial_{\alpha})A_{\mu}=0
\end{equation}
We have used the mostly negative metric : $\eta_{\mu\nu}=diag (1,-1,-1)$.  Now using a plane wave solution for $A_{\mu}$ i.e. $A_{\mu}=\chi_\mu(k)e^{-ik.x}$,  $\chi_{\mu}(k)$ being the polarisation vector, we get, 
\begin{equation}
\label{eq:5}
    (k^2\eta^{\mu\beta}-i\theta \epsilon^{\mu\alpha\beta}k_{\alpha})\chi_{\mu}(k)=0
\end{equation}
 Non-trivial solution for $\chi_{\mu}(k)$ exists when 
\begin{equation}
    \det(k^2\eta^{\mu\beta}-i\theta \epsilon^{\mu\alpha\beta}k_{\alpha})=0\\
    \implies k^4(k^2-\theta^2)=0
\end{equation}
Thus, the solutions are $k^2=0$ and $k^2=\theta^2$. For $k^2=0$, The solution for the polarisation vector is of the form $\chi_{\mu}(k)= f(k) k_{\mu}$, where $f(k)$ is an arbitrary function of $k$. So only the massive case i.e. $k^2=\theta^2$ can be taken into consideration because the former solution renders the vector $A_\mu(x)$ to be in the pue gauge and thereby is of no significance.\\
Now considering the massive case, we can get to the rest frame and write $k^{\mu}=\begin{pmatrix}|\theta|\\0\\0\end{pmatrix}$ in \eqref{eq:5}, to get
\begin{equation}
     \label{eq:7}
    \chi_0(\overrightarrow{0})=0
\end{equation}
and \eqref{eq:5} yields, for the value of $\beta=1,2$ 
\begin{equation}
\label{eq:8}
  \chi^2(\vec{0})= \mp i\chi^1(\vec{0})
\end{equation}
With this we get two linearly independent polarization vectors in rest frame and are given by \cite{1}

\begin{equation}
    \label{eq:11}
    \chi^\mu_\pm(\overrightarrow{0})=\frac{1}{\sqrt{2}}\begin{pmatrix}0\\1\\\pm i\end{pmatrix}
\end{equation}
where we have imposed the normalisation condition on $\chi^\mu$ as,
\begin{equation*}
    {\chi^\mu}^\dagger_\lambda(\overrightarrow{0})\chi_{\mu,\lambda}(\overrightarrow{0})=-1\quad;\quad\lambda=+,-
\end{equation*}
In contrast to Maxwell field theory in 2+1 dimension \cite{1} , here the polarisation vector is complex and reflect that the quanta has two transverse degrees of freedom. Note that argument $\overrightarrow{0}$ here indicates the vanishing of two spatial components of $k^\mu=(\theta,\overrightarrow{0})$ in the rest frame.
We can check the orthonormality condition given by
\begin{equation}
\label{a5}
\chi^{*\mu}_{\,(\lambda)}(k)\chi_{\mu\,(\lambda ')}(k)=\delta_{\lambda \lambda '};\,\,\,\,\lambda,\lambda ' =+,-
\end{equation}
Also the completeness relation for the polarization vectors is given by
\begin{equation}
\label{a6}
\sum _{\lambda=1,2} \chi_{\mu\,(\lambda)}(k)\,\,\chi^*_{\,\nu\,(\lambda)}(k)= \eta_{\mu\nu}-\frac{k_{\mu}k_{\nu}}{k^2}
\end{equation}
in an arbitrary Lorentz frame, where the polarisation vector for  a general Lorentz frame $\chi^\mu(k)$ can be obtained \cite{14} by performing a simple Lorentz transformation of $\chi^{\mu}(\overrightarrow{0})$ as,
\begin{equation*}
    \chi^\mu(k)={\Lambda^\mu}_\nu\chi^\nu(\overrightarrow{0})
\end{equation*}
where ${\Lambda^\mu}_\nu$ is the  Lorentz transformation matrix.  We can verify the validity of \eqref{a5} and \eqref{a6} quite easily by going to the rest frame.\footnote{It can be argued simply as: \eqref{a5} being a Lorentz invariant quantity obviously persists to be true in any Lorentz frame . And the right hand side of  \eqref{a6} is a tensor under Lorentz transformation. So if the relation \eqref{a6} holds for rest frame, it continues to hold for any boosted frame. }.\\
We are now in a position to write down the field decomposition of $A_{\mu}(x)$ as \cite{14}:
\begin{equation}
    \label{eq:15}
    A_\mu(x)=\int\frac{d^2k}{(2\pi)^22k_0}\sum\limits_{\lambda=1,2}[\chi^{(\lambda)}_\mu(k)a^{(\lambda)}(k)e^{-ik.x}+\chi^{(\lambda)}_\mu(k){a^{(\lambda)}}^\dagger(k)e^{ik.x}]
\end{equation}

\section{The probability distribution}
In this section we want to address the main motive of the paper i.e. the investigation of emission or absorption of the massive quanta of the MCS theory in presence of a classical conserved current.\\
It is quite well known that (see for example in \cite{7}) the interaction with an external current gives rise to photonic emission from vacuum and the distribution of photon number is Poisson distribution. When the dimension of the space-time is reduced to 2+1, then the photons become spinless. As reviewed briefly in section 2, the inclusion of a topological term indeed introduces one extra degree of freedom to the system thereby rendering the quanta massive. Now we can include an external source term to the MCS Lagrangian and calculate the emission distribution. The interaction of the field with a classical external current $J^{\mu}$  is given by $A_{\mu}J^{\mu}$ so that the total Lagrangian is given by,
\begin{equation}
\label{eq:a2}
 \mathscr{L}=-\frac{1}{4}F^{\mu\nu}F_{\mu\nu}+\frac{\theta}{2}\epsilon^{\mu\nu\rho}A_{\mu}\partial_{\nu}A_{\rho}+A_{\mu}J^{\mu}
\end{equation}
The solution of this equation is a simple modification of \eqref{eq:4} i.e.
\begin{equation}
\label{eq:a3}
  (-\partial^2\eta^{\mu\beta}+\theta \epsilon^{\mu\alpha\beta}\partial_{\alpha})A_{\mu}=J^{\beta}
\end{equation}
For this solution to exist in the same Fock space as of the free field, there should exist a canonical transformation relating the interacting field with the free field. Here the source $J^{\mu}$ is considered to be switched on and off adiabatically so that the system remains in the eigen state of the Hamiltonian \cite{5}. Let us refer the free fields before and after turning the source on and off respectively as $A_{\mu}^{in}$ and $A_{\mu}^{out}$.  Now the general solution of the field can be written as,
\begin{align}
\label{eq:a4}
A_{\mu}(x)&=A_{\mu}^{in}(x)+\int  d^3y\,\, G_{R}(x-y)J_{\mu}(y)\nonumber\\
&=A_{\mu}^{out}(x)+\int d^3y\,\,G_{A}(x-y)J_{\mu}(y)
\end{align}

So we can write 
\begin{equation}
\label{eq:a5}
A_{\mu}^{out}(x)=A_{\mu}^{in}(x)+\int d^3y [G_{R}-G_{A}](x-y) J_{\mu}(y)
\end{equation}
where $G_A$ and $G_R$ are advanced and retarded Green's functions. \\
It can be easily proved from the expression of free field \eqref{eq:15} that the non-equal time commutator between the fields is given by,
 \begin{equation}
\label{eq:30}
 \begin{split}
    [A_{\mu}(x),A_\nu(y)]&=\int\frac{d^2k}{(2\pi)^22k_0}\frac{d^2k'}{(2\pi)^22k_0}\sum_{\alpha,\alpha'}\chi^{(\alpha)}_\mu(k)\chi^{(\alpha')}_\nu(k')[{a^{(\alpha)}}^\dagger(k),a^{(\alpha')}(k')]e^{-i(k.x-k'.y)}\\&=G_R(x-y)-G_A(x-y)\\&= i \eta_{\mu\nu}G(x-y) 
 \end{split}
\end{equation}
$G(x-y)$ is the so called Feynman's propagator. 
With this, \eqref{eq:a5} can be rewritten as,
\begin{equation}
\label{eq:a6}
A_{\mu}^{out}(x)=A_{\mu}^{in}(x)-i\int d^3y [A_{\mu}(x),A_{\nu}(y)]J^{\nu}(y)
\end{equation}
On the other hand the canonical transformation which connects the in and out field can be represented by unitary operator S such that
$$A^{\mu}_{out}(x)=S^{-1}A^{\mu}_{in}(x)S\quad ;|out\rangle=S^{-1}|in\rangle=S^{\dagger}|in\rangle ; \quad|in\rangle=S|out\rangle$$
So the probability amplitude for no emission of photon is given by,
\begin{equation}
\label{eq:a8}
\langle0,out|0,in\rangle=\langle0,in|S|0,in\rangle =\langle 0,out|S|0,out\rangle 
\end{equation}
Using Hadamard identity i.e. $e^ABe^{-A}=B+[A,B]+\frac{1}{2!}[A,[A,B]]+...$, we can identify  from \eqref{eq:a6} that, 
\begin{equation}
\label{eq:a7}
A_{\mu}^{out}=S^{-1} A_{\mu}^{in}S=  e^{-i\int d^3y A_{\nu}(y)J^{\nu}(y)} A_{\mu}^{in}  e^{i\int d^3y A_{\nu}(y)J^{\nu}(y)}
\end{equation}
where $S= e^{-i\int d^3y A_{\nu}(y)J^{\nu}(y)}$  is the unitary scattering matrix responsible for the in between interaction and resultant transition between in and out state of the free field.
\subsection{Calculation of the propagator}

Now taking a suitable ansatz for $A_{\mu}$ in the solution \eqref{eq:a3} we
can trivally  show that the propagator of the theory in its Fourier space is given by, 

\begin{equation}
    \Tilde{G}^{\mu\nu}(k)=\frac{k^2\eta^{\mu\nu}-k^\mu k^\nu-i\theta\epsilon^{\mu\nu\alpha}k_{\alpha}}{k^2(k^2-\theta^2)}
\end{equation}
The term containing $k^{\mu}k^{\nu}$ can be dropped  as the external current is conserved and because of that this term will not have any contribution to our calculation which will evident as we proceed\footnote{See eq \eqref{eq:39}.}. So the propagator takes the following effective form:
\begin{align*}
    \Tilde{G}^{\mu\nu}(k)=\frac{k^2\eta^{\mu\nu}-i\theta\epsilon^{\mu\nu\alpha}k_\alpha}{k^2(k^2-\theta^2)}
\end{align*}
Now we can find the fourier transformation of the above expression to give the spatial representation of the propagator as :
\begin{equation}
\label{eq:32}
    \therefore G^{\mu\nu}(x-y)=\eta^{\mu\nu}\int\frac{d^3k}{(2\pi)^3}e^{-ik.(x-y)}\frac{1}{k^2-\theta^2}+\frac{1}{\theta}\int\frac{d^3k}{(2\pi)^3}e^{-ik.(x-y)}\bigg[\frac{1}{k^2}-\frac{1}{k^2-\theta^2}\bigg]\epsilon^{\mu\nu\alpha}(ik_\alpha)
\end{equation}
Let 
\begin{equation*}
    G_{KG}(x-y)=\int\frac{d^3k}{(2\pi)^3}e^{-ik.(x-y)}\frac{1}{k^2-\theta^2}
\end{equation*}
which is the Klein-Gordon like Green function in 2+1 dimensions. And,
\begin{equation*}
    G_{EM}(x-y)=\int\frac{d^3k}{(2\pi)^3}e^{-ik.(x-y)}\frac{1}{k^2}
\end{equation*}
is the electromagnetic Green function in 2+1 dimensions. So,
\begin{equation}
\label{eq:33}
    G^{\mu\nu}_{(-)}(x-y)=g^{\mu\nu}G^{KG}_{(-)}(x-y)+\frac{\theta}{\theta^2}\big[G^{EM}_{(-)}(x-y)-G^{KG}_{(-)}(x-y)\big]\epsilon^{\mu\nu\alpha}\partial_\alpha
\end{equation}
Now,
\begin{equation}
\label{eq:34}
    G^{EM}_{(-)}(x-y)=\frac{1}{(2\pi)^2}\int d^3ke^{-ik.(x-y)}\theta(k_0)\delta(k^2)
\end{equation}
And,
\begin{equation}
\label{eq:35}
    G^{KG}_{(-)}(x-y)=\frac{1}{(2\pi)^2}\int d^3ke^{-ik.(x-y)}\theta(k_0)\delta(k^2-\theta^2)
\end{equation}
\subsection{S Matrix and probability distribution}
 Here on-wards we move in the direction of calculating the probability distribution  for the Maxwell-Chern-Simons field. From \eqref{eq:15} the S-matrix as shown in \eqref{eq:a7} can be written as
\begin{equation}
\label{eq:36}
\begin{split}
    S&=\exp{\Bigg[(-i)\int A^\mu(x) J_\mu(x) d^3x\Bigg]}\\&=\exp{\Bigg[(-i)\int A^{(-)\mu}(x) J_\mu(x) d^3x\Bigg]}\exp{\Bigg[(-i)\int A^{(+)\mu}(x) J_{\mu}(x)d^3x\Bigg]}\\&\exp{\Bigg[-\frac{1}{2}\int\int d^3xd^3y[A^{(-)\mu}(x)J_{\mu}(x),A^{(+)\nu}(y)J_{\nu}(y)]\Bigg]}
\end{split}
\end{equation}
where, $A^{(-)}_{\mu}(x)$ and $A^{(+)}_{\mu}(x)$ are negative and positive frequency part of the field decomposition and
\begin{equation}
    \label{eq:37}
    J^\mu(k)=\int d^3xJ^{\mu}(x)e^{-ik.x}
\end{equation}
Considering $J^\mu$ to be real,
\begin{equation}
   \label{eq:38}
   \therefore J^{\mu}(-k)={J^{\mu}}^*(k)
\end{equation}
Now, using \eqref{eq:30} and \eqref{eq:37} the last term of \eqref{eq:36} can be evaluated as
\begin{equation}
\label{eq:39}
\begin{split}
    \int\int d^3xd^3y[A^{(-)}(x)J(x),A^{(+)}(y)J(y)]&=\int\int d^xd^3y\big[A^{(+)\mu}(x),A^{(-)\nu}(y)\big]J_\mu(x)J_\nu(y)\\&=\int\int d^3xd^3yG^{\mu\nu}_{(-)}(x-y)J_\mu(x)J_\nu(y)
\end{split}
\end{equation}
From \eqref{eq:33},
\begin{align}
\label{eq:40}
    &\int\int d^3xd^3y[A^{(-)}(x)J(x),A^{(+)}(y)J(y)]\nonumber\\&=\int\int d^3xd^3y\bigg\{\eta^{\mu\nu}G^{KG}_{(-)}(x-y)J_{\mu}(x)J_\nu(y)+\frac{\theta}{\theta^2}\bigg[G^{E\theta}_{(-)}(x-y)-G^{KG}_{(-)}(x-y)\bigg]\epsilon^{\mu\nu\alpha}\partial_\alpha\big(J_\mu(x)J_\nu(y)\big)\bigg\}
\end{align}
From \eqref{eq:34}, \eqref{eq:35} and \eqref{eq:38},
\begin{align}
\label{eq:42}
    \int\int d^3xd^3y[A^{(-)}(x)J(x),A^{(+)}(y)J(y)]=\frac{1}{(2\pi)^2}&\int d^3k\theta(k_0)\Bigg\{\delta(k^2-\theta^2)J^*(k).J(k)+\frac{\theta}{\theta^2}\bigg[\delta(k^2)-\delta(k^2-\theta^2)\bigg]\nonumber\\&\epsilon^{\mu\nu\alpha}e^{-ik.(x-y)}\int d^3x\int d^3y\big(J_\mu(x)\partial_\alpha J_\nu(y)\big)\Bigg\}
\end{align}
According to Helmholtz's theorem $J^{\mu}(k)$ can be splitted into transverse and longitudinal components as follows.
\begin{equation*}
    J^{\mu}(k)=k^{\mu}J_l(k)+J^{\mu}_{tr}(k)
\end{equation*}
where $J_l(k)$ is a number and $J^{\mu}_{tr}(k)$ is a vector orthogonal to $k$. Hence,
\begin{equation}
    {J^{\mu}}^*J_{\mu}=k^2(J_l(k))^2+|J^{\mu}_{tr}(k)|^2
\end{equation}
For a massive field,
\begin{equation*}
    {J^{\mu}}^*J_{\mu}=\theta^2(J_l(k))^2+|J^{\mu}_{tr}(k)|^2
\end{equation*}
Also, from the conservation of the current it turns out that $J_l(k)=0$. So,
\begin{equation}
    \label{eq:44}
    {J^{\mu}}^*J_{\mu}=|J^{\mu}_{tr}(k)|^2
\end{equation}
Now on putting \eqref{eq:44} in \eqref{eq:42} and in turn putting it in \eqref{eq:36},
\begin{align}
    \label{eq:45}
    S=\exp  \{&\Bigg[(-i)\int A^{(-)}_\mu J^\mu d^3x\Bigg]\}\exp\{\Bigg[(-i)\int A^{(+)}_\mu J^{\mu}d^3x\Bigg]\}\nonumber\\& \exp\{\Bigg[-\frac{1}{2}\frac{1}{(2\pi)^2}\int d^3k\theta(k_0)\Bigg\}\{\delta(k^2-\theta^2)|J^{\mu}_{tr}(k)|^2+\frac{\theta}{\theta^2}\bigg[\delta(k^2)-\delta(k^2-\theta^2)\bigg]\nonumber\\& \epsilon^{\mu\nu\alpha}e^{-ik.(x-y)}\int d^3x\int d^3y\big(J_\mu(x)\partial_\alpha J_\nu(y)\big)\Bigg\}\Bigg]\}
\end{align}
The probability of emission in the ground state \eqref{eq:a8}  is given by
\begin{equation}
    \label{eq:46}
    p_0=|\braket{0_{out}|0_{in}}|^2=|\bra{0_{in}}S\ket{0_{in}}|^2
\end{equation}
So, embedding \eqref{eq:45} in \eqref{eq:46},
\begin{align}
    p_0=\exp\{&\Bigg[-\frac{1}{2}\frac{1}{(2\pi)^2}\int d^3k\theta(k_0)\Bigg\{&\delta(k^2-\theta^2)|J^{\mu}_{tr}(k)|^2+\frac{1}{\theta}\bigg[\delta(k^2)-\delta(k^2-\theta^2)\bigg]\nonumber\\&\epsilon^{\mu\nu\alpha}e^{-ik.(x-y)}\int d^3x\int d^3y\big(J_\mu(x)\partial_\alpha J_\nu(y)\big)\Bigg\}\Bigg]\}
\end{align}
The above equation is obtained by inserting the expressions of $A^{(-)}_\mu$ and $A^{(+)}_\mu$ in terms of the ladder operators. The last term remains because it's a c-number. The projection operator, owing to the indistinguishability of photons, over $n$ states can be written as
\begin{equation}
    P_n=\frac{1}{n!}\int\frac{d^2k_1}{(2\pi)^22k_1^0}....\frac{d^2k_n}{(2\pi)^22k_n^0}\sum_{\lambda=1,2}\ket{k_1\lambda_1,...,k_n\lambda_n}\bra{k_1\lambda_1,...,k_n\lambda_n}
\end{equation}
So probability of $n$ states can be expressed as
\begin{equation}
\label{eq:47}
    \begin{split}
        p_n&=\bra{0_{in}}P_n\ket{0_{in}}\\&=\frac{1}{n!}p_0^2
        \times\bra{0_{in}}\exp{\Bigg[(i)\int d^3xA^{(+)}_{in}(x).J(x)\Bigg]}\ket{k_1\lambda_1,...,k_n\lambda_1,in}\\&\bra{k_1\lambda_1,...,k_n\lambda_1,in}\exp{\Bigg[(-i)\int d^3xA^{(-)}_{in}(x).J(x)\Bigg]}\ket{0_{in}}
    \end{split}
\end{equation}
 Now,
\begin{equation}
    \label{eq:48}
    \int d^3xA^{(-)}_{in}(x).J(x)=\int\frac{d^2k}{(2\pi)^22k_0}\sum_{\lambda=1,2}{a^{(\lambda)}}^\dagger(k)J_{\lambda}(k)
\end{equation}
where $J_{\lambda}(k)=\chi^{(\lambda)}_{\mu}(k)J^\mu(k)$. On substituting \eqref{eq:48} in \eqref{eq:47}, the term with $n$ creation operators is
\begin{equation}
    \label{eq:49}
    \begin{split}
        &\frac{(-i)^n}{n!}\bra{k_1\lambda_1,...,k_n\lambda_1,in}\int\frac{d^2k_1'...d^2k_n'}{(2\pi)^{2n}2^nk_0^n}\sum_{\lambda=1,2}{a^{(\lambda_1')}}^\dagger(k_1')...{a^{(\lambda_n')}}^\dagger(k_n')J_{\lambda_1'}(k_1')...J_{\lambda_n'}(k_n')\ket{0_{in}}\\&=\bigg[(-i)\int\frac{d^2k}{(2\pi)^22k_0}\sum_{\lambda=1,2}J_{\lambda}(k)\bigg]^n
    \end{split}
\end{equation}
The complex conjugate transpose of \eqref{eq:49} will result to
\begin{equation}
        \label{eq:50}
        \bigg[(i)\int\frac{d^2k}{(2\pi)^22k_0}\sum_{\lambda=1,2}J^*_{\lambda}(k)\bigg]^n
\end{equation}
In order to calculate $p_n$ \eqref{eq:49} and \eqref{eq:50} are used.
\begin{align}
\label{eq:51}
    \therefore p_n=\frac{1}{n!}&\Bigg[\int\frac{d^2k}{(2\pi)^22k_0}|J^{\mu}_{tr}(k)|^2\Bigg]^n\exp\{\Bigg[-\frac{1}{(2\pi)^2}\int d^3k\theta(k_0)\Bigg\{\delta(k^2-\theta^2)|J^{\mu}_{tr}(k)|^2+\nonumber\\&\frac{1}{\theta}\bigg[\delta(k^2)-\delta(k^2-\theta^2)\bigg]\epsilon^{\mu\nu\alpha}e^{-ik.(x-y)}\int d^3x\int d^3y\big(J_\mu(x)\partial_\alpha J_\nu(y)\big)\Bigg\}\Bigg]\}
\end{align}

\section{Discussion}
The probability distribution obtained for MCS theory is much different from that of the massless Maxwell theory which has a Poissonian distribution. So, we expect to get the distribution of Maxwell theory on substituting $\theta=0$ in \eqref{eq:51}. But this not what actually happens. We get an indeterminate $\frac{0}{0}$ term in the exponential. This problem can be resolved if we get rid of this term by choosing $\partial_\alpha J_\nu(y)=0$. Upon doing so we get the probability distribution to be simply a Poissonian one, exactly same as that of Maxwell theory.
\begin{equation}
    p_n=e^{-r}\frac{r^n}{n!}
\end{equation}
where,
\begin{equation}
    r=\int\frac{d^2k}{(2\pi)^22k_0}|J^\mu_{tr}(k)|^2
\end{equation}
Hence, it can be inferred from this that MCS theory can be coupled to an external current if and only if the current is independent of position.\\
The total probability can be shown to be 1, i.e., $\sum_np_n=1$ with r the average number of photons. This r also plays the role of a rate parameter \cite{16}, denoting the variation. The more the variation, more is the probability of getting higher number of states. It can be shown that the probability of getting a one state photon is the highest when the rate is 1.\\
Practically, the Chern-Simon term has no effect on the photon emission distribution. So despite being the topological mass for the theory, the CS coupling parameter does not provide an infrared cut-off. Therefore this theory is still infrared divergent.\\ \\
\textit{Acknowledgement:}
I would like to thank Dr. Biswajit Chakraborty for being a constant support throughout, Anwesha Chakraborty, Dr. Partha Nandi and Ramkumar Radhakrishnan for having extensive discussions.

\end{document}